\documentclass[10pt]{article}

\usepackage{amsmath}
\usepackage{amssymb}
\usepackage{graphics}
\usepackage{rotating}
\usepackage{cite}
\usepackage{color}
\usepackage{fancybox}
\usepackage{pstricks}
\usepackage{xcolor}
\usepackage{multicol}

\def\0{\mbox{\tiny $0$}}
\def\1{\mbox{\tiny $1$}}
\def\2{\mbox{\tiny $2$}}
\def\3{\mbox{\tiny $3$}}
\def\4{\mbox{\tiny $4$}}
\def\5{\mbox{\tiny $5$}}
\def\6{\mbox{\tiny $6$}}
\def\7{\mbox{\tiny $7$}}
\def\8{\mbox{\tiny $8$}}
\def\9{\mbox{\tiny $9$}}
\newcommand{\cd}{black!50!red!90!}
\newcommand{\cj}{black!60!green!80!}
\newcommand{\cb}{gray!15}
\newcommand{\ct}{black!30!blue}
\begin{document}
%
\thispagestyle{empty}
\setcounter{page}{0}

{\Large \bf \color{\cj}
\begin{center}
 \fbox{\colorbox{\cb}{  \color{\ct}
\begin{tabular}{c}
CLOSED FORM EXPRESSION FOR THE\\ GOOS-H\"ANCHEN LATERAL DISPLACEMENT
 \end{tabular}
 }}
\end{center}
}

\vspace*{1cm}

{\Large \bf
\begin{center}
\color{\cd} $\boldsymbol{\bullet}$
\color{\cj} Physical Review A  93, 023801-9 (2016).
\color{\cd} $\boldsymbol{\bullet}$
\end{center}
}

\vspace*{1cm}

\begin{center}
\begin{tabular}{cc}
\begin{minipage}[t]{0,55\textwidth}
\vspace*{-0.4cm}
\color{black}
{\bf  Abstract.}  The Artmann formula provides an accurate determination of the Goos-H\"anchen lateral displacement in terms of the light wavelength, refractive index and  incidence angle. In the total reflection region, this formula is widely used in the literature and confirmed by experiments. Nevertheless, for incidence at critical angle, it tends to infinity and numerical calculations are needed to reproduce the experimental data.  In this paper, we overcome the divergence problem at critical angle  and find, for Gaussian beams,  a closed formula  in terms of modified Bessel functions of the first kind. The  formula  is in excellent agreement with numerical calculations and reproduces, for incidence angles greater than critical ones, the Artmann formula. The closed form also allows one
to understand how the breaking of symmetry in the angular distribution is responsible for the difference between  measurements done by considering  the  maximum and the mean value of the beam intensity.  The results obtained in this study clearly show the Goos-H\"anchen lateral displacement dependence  on the angular distribution shape of the incoming beam. Finally, we also present a brief comparison with experimental data and other analytical formulas found in the literature.
\end{minipage}
&
{\color{\cj} \fbox{\hspace*{-0.12cm} \color{black} {\colorbox{\cb}{
\begin{minipage}[t]{0,4\textwidth}
{\bf Manoel P. Ara\'ujo}\\
Institute of  Physics ``Gleb Wataghin''\\
State University of Campinas (Brazil)\\
{\bf \color{\ct} mparaujo@ifi.unicamp.br}
\hrule
\vspace*{0.15cm}
{\bf Stefano De Leo}\\
Department of Applied Mathematics\\
State University of Campinas (Brazil)\\
{\bf \color{\ct} deleo@ime.unicamp.br}
\hrule
\vspace*{0.15cm}
{\bf Gabriel G. Maia}\\
Institute of  Physics ``Gleb Wataghin''  \\
State University of Campinas (Brazil)\\
{\bf \color{\ct} ggm11@ifi.unicamp.br}
\end{minipage}
}}}}
\end{tabular}
\end{center}

\vspace*{1cm}

\begin{center}
{\color{\ct}
{\bf
\begin{tabular}{ll}
I. & INTRODUCTION \\
II.  & CLOSED FORM EXPRESSION FOR GAUSSIAN BEAMS\\
II. & BREAKING OF SYMMETRY AND  MEAN VALUE ANALYSIS\\
IV. & DEPENDENCE ON THE SHAPE OF THE ANGULAR DISTRIBUTION\\
V. & CONCLUSIONS\\
& \\
& \,[\,13 pages, 3 figures\,]
\end{tabular}
}}
\end{center}

\vspace*{1.4cm}

{\Large
\begin{flushright}
\color{\cd} \fbox{\hspace*{-.2cm}
\colorbox{\cb}{
\,\color{\cj}$\boldsymbol{\Sigma}$
\color{\cd}$\boldsymbol{\delta}$
\hspace*{-.1cm}\color{\cj}$\boldsymbol{\Lambda}$
}\hspace*{-.2cm}
}
\end{flushright}
}


\newpage


\section*{\large \color{\ct} I. INTRODUCTION}

When light is totally reflected, evanescent waves appear in the medium of lower refraction index\cite{B1,B2}. In this case, the interference between the incident and the reflected wave is responsible for the lateral displacement of the reflected light. This displacement is called Goos-H\"anchen (GH) shift in honor of the physicists who in 1947\cite{GH47} proposed and realized, for  transverse electric (TE) waves,  an experiment to show this phenomenon.  They were able to observe  this lateral shift by a multiple reflection experimental device. One year later\cite{Art}, in order to theoretically explain the results obtained in the GH experiment, Artmann   proposed an analytical formula based on the phase difference between the incident and the reflected beam. His formula was in excellent agreement with the experimental data obtained by Goos and H\"anchen  and, more important, the new formula predicted a different shift for transverse magnetic (TM) waves. In 1949\cite{GH49}, new measurements done by Goos and H\"anchen confirmed the polarization dependence suggested by Artmann.

To make our presentation self-contained, let us take a look at the Artmann derivation  dating back to 1949.  For an optical beam propagating through the dielectric layout drawn in Fig.\,1a, the transversal displacement ($d_{_{\rm GH}}$)  can be expressed in terms of
the Artmann shift ($\delta_{_{\rm GH}}$) by an appropriate geometrical factor,
\begin{equation}
d_{_{\rm GH}} =  \frac{\cos\varphi_{\0}\,\cos \theta_{\0}}{\cos\psi_{\0}}\,\,\delta_{_{\rm GH}}\,\,.
\end{equation}
For a  multiple block dielectric structure, composed for example by $N$ triangular prisms, the final displacement will be $N\,d_{_{\rm GH}}$; see Fig.\,1(c). The Fresnel coefficients of the reflected waves at the down dielectric-air interface of the triangular prism of Fig.\,1(a) are given by
\begin{equation}
\left\{\,R^{^{\rm [TE]}} \,,\,\, R^{^{\rm [TM]}}\,\right\} = \left\{\,\frac{n\,\cos\varphi\, - \sqrt{1-n^{^{2}}\sin^{\2}\varphi}}{n\,\cos\varphi\, + \sqrt{1-n^{^{2}}\sin^{\2}\varphi}}\,\,,\,\, \frac{\cos\varphi\, -n\, \sqrt{1-n^{^{2}}\sin^{\2}\varphi}}{\cos\varphi\, + n\,\sqrt{1-n^{^{2}}\sin^{\2}\varphi}}  \,  \right\}\,\,.
\end{equation}
These coefficients gain an additional  phase when $n \sin \varphi >1$.  This  phase is different for TE and TM waves,
\begin{equation}
\left\{\,\phi_{_{\rm GH}}^{^{\rm [TE]}}\,,\,\phi_{_{\rm GH}}^{^{\rm [TM]}}\,\right\}
= -\,2\,\left\{\,\arctan\left[\,\frac{\sqrt{n^{\2}\sin^{\2}\varphi-1}}{n\cos\varphi}\,\right]\,,\,
\arctan\left[\,\frac{n\,\sqrt{n^{\2}\sin^{\2}\varphi-1}}{\cos\varphi}\,\right]
\,\right\}\,\,.
\end{equation}
The first-order term in the Taylor expansion of these new phases is responsible for the lateral displacements\cite{SPM}. In order  to quantify these shifts, Artmann used the stationary condition,
\begin{equation}
\left\{\,\frac{\partial}{\partial\varphi}\,\left[\,  n\,k\,\left(y_*\sin\varphi - z_* \cos\varphi\right) +
\phi_{_{{\rm GH}}}^{^{\rm [TE,TM]}}  \,\right]\,\right\}_{\0}=0
\end{equation}
$(k=2\,\pi/\lambda)$, obtaining, for the reflected beam intensity, a maximum which moves along
\begin{equation*}
y_{*}=    -\,z_*\,\tan\varphi_{\0} + \,\delta_{_{{\rm GH\,(Art)}}}^{^{\rm [TE,TM]}}\,\,,
\end{equation*}
with
\begin{eqnarray}
\left\{\, \delta_{_{{\rm GH\, (Art)}}}^{^{\rm [TE]}}\,,\,\,\delta_{_{{\rm GH\, (Art)}}}^{^{\rm [TM]}} \,\right\}
& = & -\, \frac{1}{n\,k\,\cos\varphi_{\0}}\,\,\left\{\,\frac{\partial \phi_{_{{\rm GH}}}^{^{\rm [TE]}}  }{\partial\varphi}\,,\,\frac{\partial \phi_{_{{\rm GH}}}^{^{\rm [TM]}}  }{\partial\varphi}
\,\right\}_{\0}
\nonumber \\
 & = &
\frac{2\tan\varphi_{\0}}{k\,\sqrt{n^{^2}\sin^{\2}\varphi_{\0}-1}}
\left\{\,1\,,\,\frac{1}{n^{^2}\sin^{\2}\varphi_{\0}-\cos^{\2}\varphi_{\0}}\,\right\}\,\,.
\end{eqnarray}
The different transversal shift for TE and TM waves  was predicted by Artmann in 1949\cite{Art}. The experimentally observed values,
\begin{eqnarray}
\label{condS}
\left\{\, d_{_{{\rm GH\,(Art)}}}^{^{\rm [TE]}}\,,\,\,d_{_{{\rm GH\,(Art)}}}^{^{\rm [TM]}} \,\right\} &=&  \frac{\cos\theta_{\0}\,\cos\varphi_{\0}}{\cos\psi_{\0}}\,
\left\{\, \delta_{_{{\rm GH\,(Art)}}}^{^{\rm [TE]}}\,,\,\,\delta_{_{{\rm GH\,(Art)}}}^{^{\rm [TM]}} \,\right\}
\nonumber \\
 & = &  -\,\frac{1}{k}\, \left\{\,\frac{\partial \phi_{_{{\rm GH}}}^{^{\rm [TE]}}  }{\partial\theta}\,,\,\frac{\partial \phi_{_{{\rm GH}}}^{^{\rm [TM]}}  }{\partial\theta}
\,\right\}_{\0}\,\,,
\end{eqnarray}
confirmed Artmann's prediction for incidence far enough from the critical region\cite{GH47,GH49,Exp1,Exp2,Exp3,Exp4}.

The question now is how to remove the infinity at critical angles,
\[ \varphi_c=\arcsin[1/n]\,\,\,\,\,\,\,\Rightarrow\,\,\,\,\,\,\,
\theta_c=\arcsin[\,(1-\sqrt{n^{^{2}}-1}\,\,)\,/\,\sqrt{2}\,]\,\,,\]
and, consequently, the discrepancy between the Artmann formula and the experimental data in the critical region\cite{Exp3}. Notwithstanding the great interest in the  literature\cite{GH1,GH2,GH3,GH4,GH5} (for clear and detailed reviews on the GH shift see Refs.\,\cite{Rev1,Rev2}), the divergence problem of the Artmann formula was  often overcome by using numerical calculations\cite{GH4,Num1,Num2} to fit the experimental curves\cite{Exp1,Exp2,Exp3,Exp4}. An analytical formula for the amplification of the GH shift at critical angle was  recently proposed\cite{GH4} but it is only valid for incidence at critical angles and does not explain the maximum obtained by numerical calculation for angles greater than the critical one.  In this paper, we aim to obtain the analytical formula for the {\em full} critical region. The possibility to have an analytical formula represents a great advantage in studying the behavior of light at the dielectric-air interface. For example, such a formula could allow  one to understand the reason for the amplification in the critical region, to calculate the incidence which guarantees the maximal shift, to study the breaking of symmetry in the angular distribution and, clearly, to avoid numerical calculations to obtain, for different refractive indices, the GH displacement plots in the critical region.

\section*{\large \color{\ct} II. CLOSED FORM EXPRESSION FOR GAUSSIAN BEAMS}

The laser  propagation can be described by assuming that the beam has an ideal Gaussian intensity profile. So, we consider an incident  beam moving in the plane $yz$ along the $z$ axis forming an angle $\theta_{\0}$ with the $\widetilde{z}$ axis, normal at the left air-dielectric interface (see Fig.\,1a),
\begin{equation}
E_{_{\rm inc}} =  E_{\0}\,\int_{_{\mbox{\footnotesize $-\,\pi/2$}}}^{^{\mbox{\footnotesize $+\,\pi/2$}}}
\hspace*{-0.7cm}\mbox{d}\theta\,\,\, g(\theta-\theta_{\0})\,\exp[\,i\,k\,(\,\widetilde{y}\,\sin\theta+\widetilde{z}\,\cos\theta\,)\,]\,\,,
 \end{equation}
with angular gaussian distribution given by
\begin{equation}
g(\theta-\theta_{\0})= \frac{k\,\mbox{w}_{\0}}{2 \,\sqrt{\pi}}\, \exp \left[-\,(\,k\,\mbox{w}_{\0}\,)^{^{2}} (\theta -\theta_{\0})^{^2}/\,4\,\right]\,\,,
\end{equation}
where $\mbox{w}_{\0}$ is the radius of the $1/e^{\2}$ irradiance contour at the plane where the wave front is flat.

The idea which suggested how to remove the infinity is very simple. The stationary phase condition given in Eq.\,(\ref{condS}) clearly represents a limit case. For example, it does not take into account the particular shape of the angular distribution which characterizes the incoming optical beam. It gives a good approximation for the GH shift when $\partial \phi_{_{\rm GH}}
/\partial \theta$ can be calculated in $\theta=\theta_{_{0}}$ and consequently factorized from the integral containing the angular distribution.  Thus, in looking for a correct generalization of the stationary condition, we have to calculate, as done for the numerical calculations which reproduce the experimental data, the following integral
\begin{equation}
d_{_{\rm GH}}^{^{\rm [TE]}}  =  -\,\frac{1}{k}\,
\int_{_{\mbox{\footnotesize $n\sin\varphi=1$}}}^{^{\mbox{\footnotesize $+\,\pi/2$}}}
\hspace*{-0.7cm}\mbox{d}\theta
\,\,\,g(\theta-\theta_{\0})\,\frac{\partial \phi_{_{\rm GH}}^{^{\rm [TE]}}}{\partial \theta\,\,}\,\mbox{\huge $/$}
 \int_{_{\mbox{\footnotesize $-\,\pi/2$}}}^{^{\mbox{\footnotesize $+\,\pi/2$}}}
\hspace*{-0.7cm}\mbox{d}\theta\,\,\, g(\theta-\theta_{\0})\,\,.
\end{equation}
Without loss of generality, for $k\,{\rm w}_{\0}\gg 1 $ and $\theta_{\0}\leq \pi/\,2 - 5/\,k{\rm w}_{\0}$, we can approximate the previous equation as follows
\begin{equation}
\label{dgh}
d_{_{\rm GH}}^{^{\rm [TE]}}
=  \frac{2\,\sin\varphi_{\0}\cos\theta_{\0}}{k\,\cos\psi_{\0}}\,
\int_{_{\mbox{\footnotesize $n\sin\varphi=1$}}}^{^{\mbox{\footnotesize $+\,\infty$}}}
\hspace*{-.5cm}\mbox{d}\theta\,\,\,\frac{g(\theta-\theta_{\0})}{\sqrt{n^{^2}\sin^{\2}\varphi-1}}
\,\,.
\end{equation}
The problem now is to analytically solve this integral. So,
let us  expand around $\theta_{\0}$ the argument of the square root which appears  in the denominator,
\begin{equation}
n^{\2}\sin^{\2}\varphi\,-\,1  \approx  n^{\2}\sin^{\2}\varphi_{\0}\,-1\, +\,   \,\frac{n\,\sin(2\,\varphi_{\0})\cos{\theta_{\0}}}{\cos\psi_{\0}}\,\,\,(\theta -\theta_{\0})
  =   \frac{n\,\sin(2\,\varphi_{\0})\cos{\theta_{\0}}}{\cos\psi_{\0}}\,\,\left(\,\theta - \theta_{\0} - \sigma_{\0}\,\right)\,\,,
\end{equation}
with
\begin{equation*}
\sigma_{\0} = \frac{\left(1- n^{\2}\sin^{\2}\varphi_{\0} \right)\,\cos\psi_{\0}}{n\,\sin(2\,\varphi_{\0})\cos{\theta_{\0}}}\,\,\,.
\end{equation*}
By using this expansion, Eq.\,(\ref{dgh}) can be rewritten as
\begin{equation}
\label{inte}
d_{_{\rm GH}}^{^{\rm [TE]}}  =
{\rm w}_{\0}\,\,\sqrt{\frac{\tan\varphi_{\0}\cos\theta_{\0}}{2\,\,n\,\pi\,\cos\psi_{\0}}}\,\,
  \int_{_{\mbox{\footnotesize $ \theta_{\0}+\sigma_{\0}$}}}^{^{\mbox{\footnotesize $+\,\infty$}}}
\hspace*{-.7cm}\mbox{d}\theta\,\,\,\,\frac{\exp \left[-\,(\,k\,\mbox{w}_{\0}\,)^{^{2}} (\theta -\theta_{\0})^{^2}/\,4\,\right]}{\sqrt{ \theta-\theta_{\0}-\sigma_{\0}}}\,\,.
\end{equation}
By introducing  the new integration variable $\rho=
k\,{\rm w}_{\0}\,\left(\,\theta - \theta_{\0} - \sigma_{\0}\right)/\,2$ and expanding the integrand in  a Taylor series, we find
\begin{eqnarray}
d_{_{\rm GH}}^{^{\rm [TE]}}
 &= &\sqrt{\frac{{\rm w}_{\0}}{k}\,\,\frac{\tan\varphi_{\0}\cos\theta_{\0}}{n\,\pi\,\cos\psi_{\0}}}\,\,
 \int_{{0}}^{{\infty}}
\hspace*{-.3cm}\mbox{d}\rho\,\exp\left[-\,\left(\rho \,+\,
\frac{k\,{\rm w}_{\0}\sigma_{\0}}{2}\right)^{^{2}}\right]\,\mbox{\huge /}\,\sqrt{\rho}\nonumber\\
 & = &  \sqrt{\frac{{\rm w}_{\0}}{k}\,\,\frac{\tan\varphi_{\0}\cos\theta_{\0}}{n\,\pi\,\cos\psi_{\0}}}\,\,    \exp\left[-\,\left(\frac{k\,{\rm w}_{\0}\sigma_{\0}}{2}\,\right)^{^{2}}\right]\,\sum_{_{m=0}}^{^{\infty}}\frac{\,\,
(-\,k\,{\rm w}_{\0}\sigma_{\0})^{^{m}}}{m!}\,
\int_{_{0}}^{^{\infty}}\hspace*{-0.3cm}\mbox{d}\rho\,\,\,e^{-\,\rho^{^2}}\,\rho^{^{m\,-\,\frac{1}{2}}} \nonumber \\
 & = & \frac{1}{2}\,\,\sqrt{\frac{{\rm w}_{\0}}{k}\,\,\frac{\tan\varphi_{\0}\cos\theta_{\0}}{n\,\pi\,\cos\psi_{\0}}}\,\,    \exp\left[-\,\left(\frac{k\,{\rm w}_{\0}\sigma_{\0}}{2}\,\right)^{^{2}}\right]\,\sum_{_{m=0}}^{^{\infty}}\frac{\,\,
(-\,k\,{\rm w}_{\0}\sigma_{\0})^{^{m}}}{m!}\,\,\Gamma\left[\frac{1\,+\,2\,m}{4}\right]\,\,.
\end{eqnarray}
It is interesting to observe that, after algebraic manipulations, the series, appearing in the previous equation, can be expressed in terms of the modified Bessel functions of first kind,
\[ I_{\alpha}(x) =\sum_{_{m=0}}^{^{\infty}}\frac{\,\,\,\,\,\,\,\,\,\,\,
(x/2)^{^{2\,m+\alpha}}}{m!\,\,\Gamma\left[m+1+\alpha\right]}\,\,. \]
Let us introduce the variable $x= k\,{\rm w}_{\0}\sigma_{\0}/2\sqrt{2}$ \,and  begin the calculation of
\[
\exp[\,-\,2\,x^{\2}]\,\sum_{_{m=0}}^{^{\infty}}\frac{( -\,2\,\sqrt{2}\,\,x )^{^{m}}}{m!}\,\Gamma\left[\frac{1\,+\,2\,m}{4}\right]\,\,.
\]
Observing that
\begin{eqnarray}
\frac{\exp[\,-\,x^{\2}]}{2^{^{1/4}} \pi}\,\sum_{_{m=0}}^{^{\infty}}\frac{( -\,2\,\sqrt{2}\,\,x )^{^{m}}}{m!}\,\Gamma\left[\frac{1\,+\,2\,m}{4}\right]  & = &\nonumber \\
 & & \hspace*{-5cm}\frac{\Gamma[\frac{1}{4}]}{2^{^{1/4}}\pi}\,\left\{\,1 + \frac{\,\,x^{^{4}}}{3}+\frac{\,\,\,x^{^{8}}}{42} + \frac{\,\,\,\,x^{^{12}}}{1386} +
... +    \frac{\sqrt{2}\,\pi}{m!\,4^{^{m}}\,\Gamma[m+\frac{3}{4}]\,\Gamma[\frac{1}{4}]}\,x^{^{4m}}     + ... \right\}
\nonumber \\
& & \hspace*{-5.8cm}-\,\frac{2^{^{1/4}}\,\Gamma[\frac{3}{4}]}{\pi}\,\left\{\,2\,x + \frac{2\,x^{^{5}}}{5}+\frac{\,\,\,x^{^{9}}}{45} + + \frac{\,\,\,\,x^{^{13}}}{1755} +
... +    \frac{\pi/\sqrt{2}}{m!\,4^{^{m}}\,\Gamma[m+\frac{5}{4}]\,\Gamma[\frac{3}{4}]}\,x^{^{4m+1}}     + ... \right\} \nonumber \\
& = & 2^{^{1/4}}\,\sum_{_{m=0}}^{^{\infty}}\frac{\,\,\,\,\,\,\,\,\,\,\,
(x/2)^{^{2\,m}}}{m!\,\,\Gamma\left[m+\frac{3}{4}\right]} - \frac{x}{2^{^{1/4}}}
\,\sum_{_{m=0}}^{^{\infty}}\frac{\,\,\,\,\,\,\,\,\,\,\,
(x/2)^{^{2\,m}}}{m!\,\,\Gamma\left[m+\frac{5}{4}\right]}
\nonumber \\
  & =  & \sqrt{|x|}\,I_{_{-1/4}}\left(\,x^{\2}\right) - \frac{x}{ \sqrt{|x|}}
  \, I_{_{1/4}}\left(\,x^{\2}\right) \,\,.
\end{eqnarray}
and introducing the new function
\begin{equation}
\mathcal{S}[\,x\,] = \exp[\,-\,x^{\2}]\,\,\sqrt{|x|}\,\left[I_{_{-1/4}}\left(\,x^{\2}\right) - {\rm sign}(x)\, I_{_{1/4}}\left(\,x^{\2}\right) \right]\,\,,
\end{equation}
we can finally give the  closed form for the GH shift in terms of the modified Bessel function of the first kind, i.e.
\begin{equation}
\label{newf}
d_{_{\rm GH}}^{^{\rm [TE]}} \, =\, \sqrt{\frac{\,\,\pi \,\tan\varphi_{\0} \cos\theta_{\0}}{2\,\sqrt{2}\,\,n\,\cos\psi_{\0}}}\,\,\mathcal{S}\left[\,\frac{k\,{\rm w}_{\0}\sigma_{\0}}{2\,\sqrt{2}}\,\right]\,\,\sqrt{\frac{{\rm w}_{\0}}{k}}\,\,.
\end{equation}
The main goal of our derivation is to remove divergence. Indeed, at a critical angle, $\sigma_{\0}=0$,   observing that
\[\lim_{x\,\to\, 0}   \mathcal{S}[\,x\,] =  \frac{\Gamma\left[\frac{1}{4}\right]}{2^{^{1/4}}\,\pi}\,\,,    \]
we now find
\begin{equation}
\label{GHcri}
d_{_{\rm GH}}^{^{\rm [TE,cri]}}   =   \sqrt{\frac{
\cos\theta_{_{\rm cri}}}{\cos\psi_{_{\rm cri}}}\,\frac{\tan\varphi_{_{\rm cri}}}{n\,\pi}\,}\,\,\,\,
\frac{\Gamma[\frac{1}{4}]}{2}\,\,\,\sqrt{\frac{{\rm w}_{\0}}{k}} \,\, \approx\, \frac{\Gamma[\frac{1}{4}]}{2\,\sqrt{n\,\pi}\,(n^{^2}-1)^{^{1/4}}}\,\,\sqrt{\frac{{\rm w}_{\0}}{k}}\,\,,
\end{equation}
where the last approximation comes from the fact that in the critical region $\cos\theta_{_{\rm cri}}\approx \cos\psi_{_{\rm cri}}$.   Eq.\,(\ref{GHcri}) clearly shows that at critical incidence we find the amplification factor $\sqrt{k\,{\rm w}_{\0}}$  with respect to the behavior proportional to $\lambda$  predicted by Artmann, Eq.\,(\ref{Alim}).

We can also estimate from which  incidence angle it is correct to use the Artmann formula. For $k\,{\rm w}_{\0}\,\sigma_{\0}<-\,2\,\pi$,  due to the presence of the Gaussian integrand $g(\theta-\theta_{\0})$,  the lower limit of integration in Eq.\,(\ref{inte})
can be changed to $-\,\infty$ and, by using
\[\lim_{x\,\to\, -\,\infty}   \sqrt{|x|}\,\mathcal{S}[\,x\,] =  \sqrt{\frac{2}{\pi}}\,\,,\]
we find
\begin{equation}
\label{Alim}
d_{_{{\rm GH}\,\, {\footnotesize \mbox{($k{\rm w}_{\0}\sigma_{\0}\ll -\,1)$}}}}^{^{\rm [TE]}}  \,\longrightarrow\,\,\,\,    \sqrt{\frac{\tan\varphi_{\0}\cos\theta_{\0}}{\sqrt{2}\,\,n\,\cos\psi_{\0}}\,\,
\frac{2\,\sqrt{2}}{k\,{\rm w}_{\0}\,(-\,\sigma_{\0})}\,\,\frac{{\rm \,w}_{\0}}{k}}\,\,=\,\,   d_{_{\rm GH\,(Art)}}^{^{\rm [TE]}}\,\,.
\end{equation}
Observing that in the critical region  $-\,\sigma_{\0} \approx n\,\delta \varphi = \delta \theta$, we obtain the incidence angle starting from which we can use the Artmann formula
\begin{equation}
 \theta_{_{0\,(\rm Art)}} \,\geq\,\, \theta_{_{\rm cri}} \,+\, \frac{2\,\pi}{k\,{\rm w}_{\0}}\,\,=\, \theta_{_{\rm cri}} \,+\, \frac{\lambda}{\,\,{\rm w}_{\0}} \,\,.
 \end{equation}
It is interesting to observe that this angle does not depend on the refractive index. It only depends on the ratio between ${\rm w}_{\0}$ and $\lambda$.  For example, for a Gaussian He-Ne laser ($\lambda=633\,{\rm nm})$ with ${\rm w}_{\0}=1\,{\rm mm}$, we find $\delta \varphi_{_{\rm Art}} \approx 0.036^{^o}$, see Figs.\,2(c) and (d).

From the new analytical formula (\ref{newf}), we can also determine the incidence angle for which the GH shift is maximized. To do it, let us observe that the maximum of $\mathcal{S}[\,x\,]$ is found at
\[x_{_{\rm max}}= -\,0.38\,\,.\]
As was done for the Artmann limit, by using $\theta_{_0}=\theta_{_{\rm cri}}+\,\delta \theta_{_{\rm max}}$ in $\sigma_{\0}$, we obtain
\[ -\,0.38\, = \frac{k\,{\rm w}_{\0}\,\sigma_{\0}}{2\,\sqrt{2}} \approx -\,\frac{k\,{\rm w}_{\0}\,n\,\delta \varphi_{_{\rm max}}}{2\,\sqrt{2}}\approx -\,\frac{k\,{\rm w}_{\0}\,\delta \theta_{_{\rm max}}}{2\,\sqrt{2}} \,\,\,\,\,\,\,\Rightarrow\,\,\,\,\,\,\,
\delta \theta_{_{\rm max}} \approx \frac{1}{\,\,k\,{\rm w}_{\0}}\,\,.
\]
This shows that, as predicted by numerical and experimental data, the maximum is {\em not} found at critical incidence but at angles greater than the critical one,
\begin{equation}
\label{eqcr}
 \theta_{_{\rm max}} = \, \theta_{_{\rm cri}} + \frac{1}{\,\,k\,{\rm w}_{\0}}\,\,.
\end{equation}
This angle is independent of the refractive index of the dielectric block.

The results obtained for the GH shift in the case of TE waves are immediately extended to TM wave by observing that
\begin{equation}
\label{newfTM}
d_{_{\rm GH}}^{^{\rm [TM]}} \, = \,
\frac{d_{_{\rm GH}}^{^{\rm [TE]}}}{n^{^2}\sin^{\2}\varphi_{\0}-\cos^{\2}\varphi_{\0}}\,\,.
\end{equation}
At critical angles, for TM waves we find  an amplification   with respect to TE waves,
\begin{equation}
d_{_{\rm GH}}^{^{\rm [TM,cri]}}  = n^{\2}\,\, d_{_{\rm GH}}^{^{\rm [TE,cri]}}\,\,.
\end{equation}
Note that the angle in which it is found the maximal shift is the {\em same} for TE and TM waves.

In Fig.\,2, we plot the analytical curves (solid blue lines) for the GH shift for TE and TM wave with $\lambda=0.633\,{\rm nm}$ and ${\rm w}_{\0}=0.5,\,1.0,\,2.0\,\,{\rm mm}$. The analytical curves show an excellent agreement with the numerical data (gray dots) obtained by calculating the GH shift of the maximum, with respect to the Snell path $\widetilde{z}=\widetilde{y}\,\tan\theta_{\0}$,  directly from the outgoing beam
\begin{equation}
E_{_{\rm out}}^{^{[{\rm TE,TM}]}} =  E_{\0}\,\int_{_{\mbox{\footnotesize $-\,\pi/2$}}}^{^{\mbox{\footnotesize $+\,\pi/2$}}}
\hspace*{-0.7cm}\mbox{d}\theta\,\,\,T^{^{[\rm TE,TM]}} g(\theta-\theta_{\0})\,\exp[\,i\,k\,(\,\widetilde{z}\,\sin\theta+\widetilde{y}\,\cos\theta\,)\,]\,\,,
 \end{equation}
with
\begin{equation}
\{\,T^{^{[\rm TE]}}\,,\,T^{^{[\rm TE]}}\,\}  = 4\,n\,\cos\theta\,\cos\psi\,
\left\{\,
\frac{R^{^{[\rm TE]}}}{(\,\cos\theta + n\,\cos\psi)^{^{2}}}
\,,\,
\frac{R^{^{[\rm TM]}}}{(\,n\,\cos\theta + \cos\psi)^{^{2}}}
\,\right\}\,\,.
\end{equation}

\section*{\large \color{\ct} III. BREAKING OF SYMMETRY AND MEAN VALUE ANALYSIS}
Once we have obtained the integral expression for  the outgoing beam, we can calculate its  mean value shift
\begin{equation}
\langle\, d_{_{\rm GH}}^{^{\rm [TE]}}\rangle   = \frac{\displaystyle{\int_{_{\mbox{\footnotesize $-\,\infty$}}}^{^{\mbox{\footnotesize $+\,\infty$}}}\hspace*{-.5cm} {\rm d} \widetilde{z} \,\,\,\,\widetilde{z}\,\cos\theta_{\0}\,\,\,\,
\left|\,E_{_{\rm out}}^{^{[{\rm TE,TM}]}}\,\right|^{^{2}}}}{\displaystyle{
\int_{_{\mbox{\footnotesize $-\,\infty$}}}^{^{\mbox{\footnotesize $+\,\infty$}}}\hspace*{-.5cm}
{\rm d} \widetilde{z} \,\,\,\,
\left|\,E_{_{\rm out}}^{^{[{\rm TE,TM}]}}\,\right|^{^{2}}}}\,\,-\,\,\widetilde{y}\,\sin\theta_{\0}\,\,.
\end{equation}
By using the $\widetilde{z}$ integration to obtain an angular $\delta$ function and using this $\delta$ function to integrate one of the angular integration variable, we get
\begin{eqnarray}
\langle\, d_{_{\rm GH}}^{^{\rm [TE]}}\rangle  & = & \frac{\displaystyle{\frac{1}{2\,i\,k}\,\int^{^{\mbox{\footnotesize $+\,\pi/2$}}}_{_{\mbox{\footnotesize $n\sin\varphi=1$}}}\hspace*{-.7cm} {\rm d} \theta \,\,\,\,\,\,\begin{array}{l}
\,\,\,\,\,g(\theta-\theta_{\0})\,\,\,\,\exp\{i\,[\,\phi_{_{\rm GH}}^{^{[\rm TE,TM]}} -\,k\,\widetilde{y}\,\sin\theta_{\0}(\theta-\theta_{\0})]\}\,\times\\
\partial_\theta\,[\,g(\theta-\theta_{\0})\,\,\,\,\exp\{i\,[\,\phi_{_{\rm GH}}^{^{[\rm TE,TM]}} -\,k\,\widetilde{y}\,\sin\theta_{\0}(\theta-\theta_{\0})]\}\,]^{^{*}}
\end{array}
}}{
\displaystyle{
\int^{^{\mbox{\footnotesize $+\,\pi/2$}}}_{_{\mbox{\footnotesize $-\,\pi/2$}}}
\hspace*{-.7cm} {\rm d} \theta
\,\,\,g^{\2}(\theta-\theta_{\0})
}
}
\,\,+\,\,{\rm h.c.}\,\,-\,\,\widetilde{y}\,\sin\theta_{\0}\nonumber \\
& \hspace*{-3cm} =  &  \hspace*{-1.5cm}    -\,\frac{1}{k}\,
 \int^{^{\mbox{\footnotesize $+\,\pi/2$}}}_{_{\mbox{\footnotesize $n\sin\varphi=1$}}}\hspace*{-.7cm} {\rm d} \theta
 \,\,\,g^{\2}(\theta-\theta_{\0})\,\frac{\partial \phi_{_{\rm GH}}^{^{\rm [TE]}}}{\partial \theta\,\,}\,\mbox{\huge $/$}
 \int^{^{\mbox{\footnotesize $+\,\pi/2$}}}_{_{\mbox{\footnotesize $-\,\pi/2$}}}
\hspace*{-.7cm} {\rm d} \theta
\,\,\,g^{\2}(\theta-\theta_{\0}) \approx
 \frac{2\,\sqrt{2\,\pi}}{k\,{\rm w}_{\0}}\,\,\,\frac{2\,\sin\varphi_{\0}\cos\theta_{\0}}{k\,\cos\psi_{\0}}\,
\int_{_{\mbox{\footnotesize $n\sin\varphi=1$}}}^{^{\mbox{\footnotesize $+\,\infty$}}}
\hspace*{-.5cm}\mbox{d}\theta\,\,\,\frac{g^{\2}(\theta-\theta_{\0})}{\sqrt{n^{^2}\sin^{\2}\varphi-1}}
\nonumber \\
& = &  \sqrt{2} \,\, {\rm w}_{\0}\,\,\sqrt{\frac{\tan\varphi_{\0}\cos\theta_{\0}}{2\,\,n\,\pi\,\cos\psi_{\0}}}\,\,
  \int_{_{\mbox{\footnotesize $ \theta_{\0}+\sigma_{\0}$}}}^{^{\mbox{\footnotesize $+\,\infty$}}}
\hspace*{-.7cm}\mbox{d}\theta\,\,\,\,\frac{\exp \left[-\,(\,k\,\mbox{w}_{\0}\,)^{^{2}} (\theta -\theta_{\0})^{^2}/\,2\,\right]}{\sqrt{ \theta-\theta_{\0}-\sigma_{\0}}}\,\,.
\end{eqnarray}
Now, without repeat the mathematical discussion presented in the previous section, we observe that this equation is obtained  by taking  the substitution
\[ {\rm w}_{\0}\,\,\,\rightarrow\,\,\,\sqrt{2}\,\,{\rm w}_{\0} \]
in Eq.(\ref{inte}).  Consequently, the result obtained in Eq.(\ref{newf}) is immediately generalized for the mean value analysis
\begin{eqnarray}
\label{newf2}
\langle\, d_{_{\rm GH}}^{^{\rm [TE]}}\rangle
 & = & \sqrt{\frac{\,\,\pi \,\tan\varphi_{\0} \cos\theta_{\0}}{2\,\sqrt{2}\,n\,\cos\psi_{\0}}}\,\,\mathcal{S}\left[\,\frac{k\,{\rm w}_{\0}\sigma_{\0}}{2}\,\right]\,\,\sqrt{\frac{\sqrt{2}\,{\rm w}_{\0}}{k}}\,\,.
\end{eqnarray}
For incidence at critical angle, we find an amplification with respect to Eq.(\ref{GHcri}),
\begin{equation}
\label{GHcri2}
\displaystyle{\langle\, d_{_{\rm GH}}^{^{\rm [TE,cri]}} \rangle  =  2^{^{1/4}}\,\,d_{_{\rm GH}}^{^{\rm [TE,cri]}}   }\,\,,
\end{equation}
and in the Artmann limit we recover the standard formula,
\begin{equation}
\label{Alim2}
\langle\,d_{_{{\rm GH}\,\, {\footnotesize \mbox{($k{\rm w}_{\0}\sigma_{\0}\ll -\,1)$}}}}^{^{\rm [TE]}} \rangle \,\,\,\,\longrightarrow\,\,\,\,    \sqrt{\frac{\tan\varphi_{\0}\cos\theta_{\0}}{\sqrt{2}\,\,n\,\cos\psi_{\0}}\,\,
\frac{2}{k\,{\rm w}_{\0}\,(-\,\sigma_{\0})}\,\,\frac{{\rm \,\sqrt{2}\,w}_{\0}}{k}}\,\,=\,\,   d_{_{\rm GH\,(Art)}}^{^{\rm [TE]}}\,\,.
\end{equation}
The relation between TE and TM waves is the same encountered in the previous section, i.e.
\begin{equation}
\label{newf2TM}
\langle\,d_{_{\rm GH}}^{^{\rm [TM]}} \rangle  \,=\,
\frac{\langle\,d_{_{\rm GH}}^{^{\rm [TE]}}\rangle}{n^{^2}\sin^{\2}\varphi_{\0}-\cos^{\2}\varphi_{\0}}\,\,.
\end{equation}
In Fig.\,2, we plot the analytical curves (red dashed lines) for the mean value GH shift for  TE and TM wave with $\lambda=0.633\,{\rm nm}$ and ${\rm w}_{\0}=0.5,\,1.0,\,2.0\,\,{\rm mm}$. The analytical curves show an excellent agreement with the numerical data (gray dots). The difference between the solid blue lines and the red dashed ones is a clear evidence of the breaking of symmetry in the angular distribution. For the incoming optical beam, due to the symmetry  of the angular distribution  the maximum and the mean values coincides. For the transmitted beam, due to the presence of $R^{^{\rm [TE,TM]}}$ in the transmission coefficient, the angular distribution is no longer  a symmetric distribution. The breaking of symmetry is clearly dependent on the beam size ${\rm w}_{\0}$, for decreasing values of ${\rm w}_{\0}$ the size of the angular Gaussian distribution
increases and consequently the breaking of symmetry caused by the transmission coefficient is more evident.

\section*{\large \color{\ct} IV. DEPENDENCE ON THE SHAPE OF THE ANGULAR DISTRIBUTION}
In the previous section, we showed the difference between maximum and mean value analyses and we saw how, for an incidence angle greater enough than the critical one to avoid the infinity, we recover the Artmann formula (\ref{condS}) which does not distinguish between the two cases.
 In the critical region, another important point comes from the dependence of the Goos-H\"anchen shift on the shape of the angular distribution. To illustrate this additional dependence, we compare the results obtained in Sec.\,II for Gaussian optical beams in the case of
an incoming beam whose spatial behavior is determined by the Fourier transform of a {\em box} angular distribution,
\begin{equation}
b(\theta-\theta_{\0})=\left\{\,\begin{array}{ccl}
\displaystyle{\frac{k\,\mbox{w}_{\0}}{2 \,\sqrt{\pi}}}   & \,\,\,\,\,  & {\rm for}\,\, \theta-\theta_{\0}
 \in\displaystyle{ \left[\,-\, \frac{\sqrt{\pi}}{k\,\mbox{w}_{\0}}\,,\,
 \frac{\sqrt{\pi}}{k\,\mbox{w}_{\0}}\,\right]\,\,,}\\ \\
0&  & {\rm otherwise}\,\,.
\end{array}\right.
\end{equation}
The choice of a box function is only because of its simple integration. Thus, what we aim to present in this section can be seen as a toy model to understand how the shape of the angular distribution influences the GH shift curves.

Following the mathematical discussion presented in Sec.\,II, we have
\begin{eqnarray}
d_{_{\rm GH,box}}^{^{\rm \,\,[TE]}} & = & -\,\frac{1}{k}\,
\int_{_{\mbox{\footnotesize $n\,\sin\varphi=1$}}}^{^{\mbox{\footnotesize $+\,\pi/2$}}}
\hspace*{-0.9cm}\mbox{d}\theta\,\,\,b(\theta-\theta_{\0})\,\frac{\partial \phi_{_{\rm GH}}^{^{\rm [TE]}}}{\partial \theta\,\,}\,\mbox{\huge $/$}
\int_{_{\mbox{\footnotesize $-\,\pi/2$}}}^{^{\mbox{\footnotesize $+\,\pi/2$}}}
\hspace*{-.7cm}\mbox{d}\theta\,\,\,b(\theta-\theta_{\0}) \nonumber \\
&= & \frac{2\,\sin\varphi_{\0}\cos\theta_{\0}}{k\,\cos\psi_{\0}}\,
\int_{_{\mbox{\footnotesize $n\,\sin\varphi=1$}}}^{^{\mbox{\footnotesize $+\,\pi/2$}}}
\hspace*{-0.9cm}
\mbox{d}\theta\,\,\,b(\theta-\theta_{\0})\,\mbox{\large /}\,\sqrt{n^{^2}\sin^{\2}\varphi-1}
\nonumber \\
&= &{\rm w}_{\0}\,\,\sqrt{\frac{\tan\varphi_{\0}\cos\theta_{\0}}{2\,\,n\,\pi\,\cos\psi_{\0}}}\,\,
  \int_{_{\mbox{\footnotesize $ \theta_{\0}+{\rm Max}[\,-\frac{\sqrt{\pi}}{k{\rm w}_{\0}}\,,\sigma_{\0}\,]$}}}^{^{\mbox{\footnotesize $ \theta_{\0}+ \frac{\sqrt{\pi}}{k{\rm w}_{\0}}$}}}
\hspace*{-2.1cm}\mbox{d}\theta\,\mbox{\large /}\,\sqrt{ \theta-\theta_{\0}-\sigma_{\0}}\nonumber \\
 &= &{\rm w}_{\0}\,\,\sqrt{\frac{2\,\tan\varphi_{\0}\cos\theta_{\0}}{n\,\pi\,\cos\psi_{\0}}}\,\,
\left(\,\sqrt{\frac{\sqrt{\pi}}{k{\rm w}_{\0}} - \sigma_{\0}  } \, -\, \sqrt{{\rm Max}\left[\,-\frac{\sqrt{\pi}}{k{\rm w}_{\0}}\,,\sigma_{\0}\,\right] - \sigma_{\0}   } \,\,\right)\,\,.
\end{eqnarray}
At critical angle, $\sigma_{\0}=0$, we find
\begin{equation}
d_{_{\rm GH,box}}^{^{\rm \,\,[TE,cri]}} =  {\rm w}_{\0}\,\,\sqrt{\frac{2\,\tan\varphi_{\0}\cos\theta_{\0}}{n\,\pi\,\cos\psi_{\0}}}\,\,
\sqrt{\frac{\sqrt{\pi}}{k{\rm w}_{\0}}} = \frac{2\sqrt{2\sqrt{\pi}}}{\Gamma[\frac{1}{4}]}\,d_{_{\rm GH,gaus}}^{^{\rm \,\,[TE,cri]}} \approx 1.04\,d_{_{\rm GH,gaus}}^{^{\rm \,\,[TE,cri]}}\,\,,
\end{equation}
and in the limit  $-\,k{\rm w}_{\0}\sigma_{\0}\gg 1$, we recover the Artmann formula
\begin{equation}
\label{Alim2}
d_{_{{\rm GH,box}\,\, {\footnotesize \mbox{($k{\rm w}_{\0}\sigma_{\0}\ll -\,1)$}}}}^{^{\rm [TE]}}  \,\longrightarrow\,\,\,\,
{\rm w}_{\0}\,\,\sqrt{\frac{2\,\tan\varphi_{\0}\cos\theta_{\0}}{n\,\pi\,\cos\psi_{\0}}}\,\frac{\sqrt{\pi}}{k{\rm w}_{\0}\sqrt{-\,\sigma_{0}}} \,\,
 \,\,=\,\,   d_{_{\rm GH\,(Art)}}^{^{\rm [TE]}}\,\,,
\end{equation}
which does not depend on the shape of the angular distribution.

In Fig.\,3, we show the difference between box (dashed green lines) and Gaussian (blue solid lines) angular distributions. The analysis of the box angular distribution  suggests that depending on the angular zone investigated we can use a simplified angular distribution to obtain a first indication on the experimental data.

\section*{\large \color{\ct} V. CONCLUSIONS}

In this paper, we have shown the validity of the GH shift's Artmann formula for incidence angles greater than
\[\theta_{_{\rm cri}}+\frac{\lambda}{\,\,{\rm w}_{\0}}\,\,\hspace*{2cm}\left[\,\varphi_{_{\rm cri}} + n\,\frac{\lambda}{\,\,{\rm w}_{\0}} \,\right]\,\,.\]
In this region, the GH shift does not depend on the shape of the angular distribution and,  due to the symmetry of the outgoing beam, does not distinguish between the maximum and the mean value intensity  measurements.

In the critical region,
\[ \theta_{_{\rm cri}}-\frac{\lambda}{\,\,{\rm w}_{\0}}\,\, \leq\,\, \theta \,\,\leq\,\, \theta_{_{\rm cri}}+\frac{\lambda}{\,\,{\rm w}_{\0}}\,\,,\]
we have found for Gaussian optical beams a {\em new} closed-form expression for the GH lateral displacement. In this region, due to the breaking of symmetry, we have to distinguish between the maximum and the mean value intensity measurements. For TE waves, we have obtained
\begin{eqnarray}
\left\{\,  d_{_{\rm GH,gaus}}^{^{\rm [TE]}}\,,\, \langle\, d_{_{\rm GH,gaus}}^{^{\rm [TE]}}\rangle\right\}
 & = & \sqrt{\frac{\,\,\pi \,\tan\varphi_{\0} \cos\theta_{\0}}{2\,\sqrt{2}\,n\,\cos\psi_{\0}}}\,\,\left\{\, \mathcal{S}\left[\,\frac{k\,{\rm w}_{\0}\sigma_{\0}}{2\,\sqrt{2}}\,\right]\,,\,2^{^{1/4}}\,
  \mathcal{S}\left[\,\frac{k\,{\rm w}_{\0}\sigma_{\0}}{2}\,\right]\,\right\}\,\,\sqrt{\frac{{\rm w}_{\0}}{k}}\,\,,
\end{eqnarray}
with the new function
\[
\mathcal{S}[\,x\,] = \exp[\,-\,x^{\2}]\,\,\sqrt{|x|}\,\left[I_{_{-1/4}}\left(\,x^{\2}\right) - {\rm sign}(x)\, I_{_{1/4}}\left(\,x^{\2}\right) \right]
\]
given in terms of the Bessel function of the first kind. The new formulas remove the divergence at the critical angle, where we now find
\begin{eqnarray}
\label{com1}
\left\{\,  d_{_{\rm GH,gaus}}^{^{\rm [TE,cri]}}\,,\, \langle\, d_{_{\rm GH,gaus}}^{^{\rm [TE,cri]}}\rangle\right\}
 & = & \frac{\Gamma[\frac{1}{4}]}{2\,\sqrt{n\,\pi}\,(n^{^2}-1)^{^{1/4}}}\frac{\,}{}\,\left\{\, 1\,,\,2^{^{1/4}}\,\right\}\,\,\sqrt{\frac{{\rm w}_{\0}}{k}}\,\,.
\end{eqnarray}
The formulas for TM wave are immediately obtained from the TE ones,
\begin{eqnarray}
\label{com2}
\left\{\,  d_{_{\rm GH,gaus}}^{^{\rm [TM]}}\,,\, \langle\, d_{_{\rm GH,gaus}}^{^{\rm [TE]}}\rangle\right\} & = &
\left\{\,  d_{_{\rm GH,gaus}}^{^{\rm [TM]}}\,,\, \langle\, d_{_{\rm GH,gaus}}^{^{\rm [TE]}}\rangle\right\}\,{\mbox{\Large /}}\,\,
\left(\,n^{\2}\sin^{\2}\varphi_{\0}-\cos^{\2}\varphi_{\0}\,\right)\,\,.
\end{eqnarray}
In 1970\cite{HT}, Horowitz and Tamir, by using a Fresnel approximation to analytically solve the integral determining the propagation of  the transmitted beam, found, for the TE and TM lateral displacement, a closed expression in terms of parabolic-cylinder (Weber) functions. In the critical region, the Horowitz-Tamir formula, translated in our notation
\[  (\,w\,,\,k_{\0}\,,\,k\,)\,\,\,\rightarrow\,\,\,(\,\mbox{w}_{\0}\,,\,k\,,\,n\,k \,) \,\,\,\,\,\,\,\,\,\,{\rm and}\,\,\,\,\,\,\,\,\,\,\theta\,\,\,\rightarrow\,\,\,\varphi_{\0}\,\,,\]
simplifies to
\begin{equation}
\delta_{_{\rm GH(HoTa)}}^{^{[\rm TE,TM]}} \approx \frac{A^{^{[\rm TE,TM]}}_{\0}}{2^{^{5/4}}\,\cos\varphi_{\0}}\,\,
\mbox{Re}\left[ e^{^{i\,\pi/4}}\,D_{_{-1/2}}\left(\,\gamma_{\0}\,\right)\right]\,\,
\exp\left(\,\gamma_{\0}^{^{2}}/4\,
\right)\,\sqrt{\frac{{\rm w}_{\0}}{n\,k}}\,\,,
\end{equation}
with
\[\gamma_{\0} = i\,\,n\,k\,{\rm w}_{\0}\,\,
\frac{\sin\varphi_{\0}-\sin\varphi_{_{\rm cri}}}{\sqrt{2}\,\cos\varphi_{\0}}\]
and
\[
\left\{\,A^{^{[\rm TE]}}_{\0}\,,\,A^{^{[\rm TM]}}_{\0}\,\right\} =
 \frac{4\,\sin\varphi_{\0}}{\sqrt{(\sin\varphi_{\0}+\sin\varphi_{_{\rm cri}})\,\cos\varphi_{_{\rm cri}}}}\,\left\{\,1\,,\, \frac{n^{\2}\,\cos^{\2}\varphi_{_{\rm cri}}}{\cos^{\2}\varphi_{\0} + n^{^4}( \sin^{\2}\varphi_{\0}- \sin^{\2}\varphi_{_{\rm cri}})}   \,\right\}\,\,.
\]
At the critical angle, $\varphi_{\0}=\varphi_{_{\rm cri}}$ implies  $\gamma_{_{\rm cri}}=\,\,0$.
Observing that
\begin{equation}
D_{_{-1/2}}\left(\,0\,\right)=  \frac{\Gamma[\frac{1}{4}]}{2^{^{3/4}}\,\sqrt{\pi}}
\,\,\,\,\,\,\,\,\,\,\,\,{\rm and}
\,\,\,\,\,\,\,\,\,\left\{\,A^{^{[\rm TE]}}_{_{\rm cri}}\,,\,A^{^{[\rm TM]}}_{_{\rm cri}}\,\right\} =\,\frac{2\,\sqrt{2}}{(n^{\2}-1)^{^{1/4}}}\,\left\{\,1\,,\,n^{\2}\,\right\}\,\,,
\end{equation}
we find
\begin{eqnarray}
\left\{\,d_{_{\rm GH(HoTa)}}^{^{[\rm TE,cri]}}\,,\,d_{_{\rm GH(HoTa)}}^{^{[\rm TM,cri]}}
 \,\right\} & = & \frac{\cos\theta_{_{\rm cri}}\cos\varphi_{_{\rm cri}}}{\cos\psi_{_{\rm cri}}}\,\, \left\{\,\delta_{_{\rm GH(HoTa)}}^{^{[\rm TE,cri]}}\,,\,\delta_{_{\rm GH(HoTa)}}^{^{[\rm TM,cri]}}
 \,\right\} \nonumber \\
& = & \frac{\Gamma[\frac{1}{4}]}{2\,\sqrt{n\,\pi}\,(n^{^2}-1)^{^{1/4}}}\,\,\left\{\, 1\,,\,n^{\2}\,\right\}\,\,\sqrt{\frac{{\rm w}_{\0}}{k}}\,\,.
\end{eqnarray}
The results obtained by Horowitz and Tamir by using the Weber functions are
in perfect agreement with our results found by using the  modified Bessel functions; see Eqs.\,(\ref{com1}) and (\ref{com2}). It is also interesting to note that the Horowitz-Tamir maximum lateral displacement is found by numerical calculations at
\begin{equation}
-\,i\,\gamma_{_{\rm max}} = \,\,0.77\,\,\,\,\,\Rightarrow\,\,\,\,\,n\,k\,{\rm w}_{\0}\,\,
\frac{\sin(\varphi_{_{\rm cri}}+\delta\varphi_{_{\rm max}})  -\sin\varphi_{_{\rm cri}}}{\sqrt{2}\,\cos(\varphi_{_{\rm cri}}+\delta\varphi_{_{\rm max}}) }\,\, = \,\,0.77\,\,\,\,\,\Rightarrow\,\,\,\,\,n\,\delta\varphi_{_{\rm max}}\approx\,\,1/\,k\,\mbox{w}_{\0}\,\,,
\end{equation}
once again in perfect agreement with the maximum displacement  given in Eq.\,(\ref{eqcr}). The formulations based on the Weber and modified Bessel functions give an analytical expression for the lateral displacement of the beam intensity maximum. In our analysis is also obtained the analytical formula for its mean value often measured in optical experiments. Clearly, starting from  Horwitz and Tamir's work it is possible to generalize their formula for the mean value case.

In an interesting paper dated 1986\cite{Cheng}, Cheng and Tang compared the Horowitz-Tamir formula with experimental data\cite{Exp2}. The Horowitz-Tamir formula as well the formula presented in this paper do not contain any axial corrections.  This means that the camera in the experiment has to be moved close to the interface from which the outgoing beam appears. Axial dependence \cite{ad06} requires a more complicated study for analytical approximations together with a weak measurement analysis \cite{wm15}. The effect of the axial dependence appear before the critical region. This effect, for example, could be responsible for the difference observed between analytical expressions and experimental data shown in Ref.\,\cite{Cheng}.

The wave-packet approach is surely essential in obtaining the GH shift. Nevertheless, other parameters also play an important  role in completing its description towards the full correspondence with experimental reality. Notably, the coherence is an important aspect of real optical systems  because real laser beams are always partially coherent. This aspect was introduced in the GH theory by Wang et al.\cite{wang08}. They investigated, by numerical calculations, the shift of the reflected beam in the presence of partial coherence. The interesting result was that as the spatial coherence of the beam decreases its GH shift reduces.
A formal expression for the GH shift of partially coherent beams in terms of the Mercer expansion recently appeared in the literature\cite{wang13}. The partial coherence as well the axial dependence have to be included in joint analytical expression with experimental data and the GH shift can be used to determine the coherence of the incoming beam \cite{wang13} as well the axial angular deviations \cite{wm15}.

The analysis done for a box angular distribution clearly shows the dependence on the distribution shape. This stimulates further investigation to obtain the analytical formula for the  GH shift
of other optical beams often used in laser experiments, such as Hermite and Laguerre beams.

We hope the study presented in this paper could be useful in understanding the effects of the breaking of symmetry in the angular distribution and in avoiding numerical calculation to test experimental data in the critical region.

\vspace{.8cm}

\noindent
\textbf{ACKNOWLEDGEMENTS}\\
The authors gratefully thank  CNPq (S.D.L. and G.G.M.) and  Faepex (M.P.A.) for the financial support. One of the authors (S.D.L.) also thanks  Silv\^ania A. Carvalho for stimulating conversations which motivated the study of an analytical expression for the GH shift in the critical region. Finally, the authors are greatly indebted to an anonymous referee for drawing their attention to the analytical work of Horowitz and Tamir\cite{HT}, the paper comparing analytical formulas with experimental data\cite{Cheng}, and the  articles showing the effect of partial coherence in the GH shift\cite{wang08,wang13}.

\newpage

\begin{figure}[ht]
	\vspace{-2.2cm}
		\hspace{-1.3cm}
		\includegraphics[height=24cm, width=18cm]{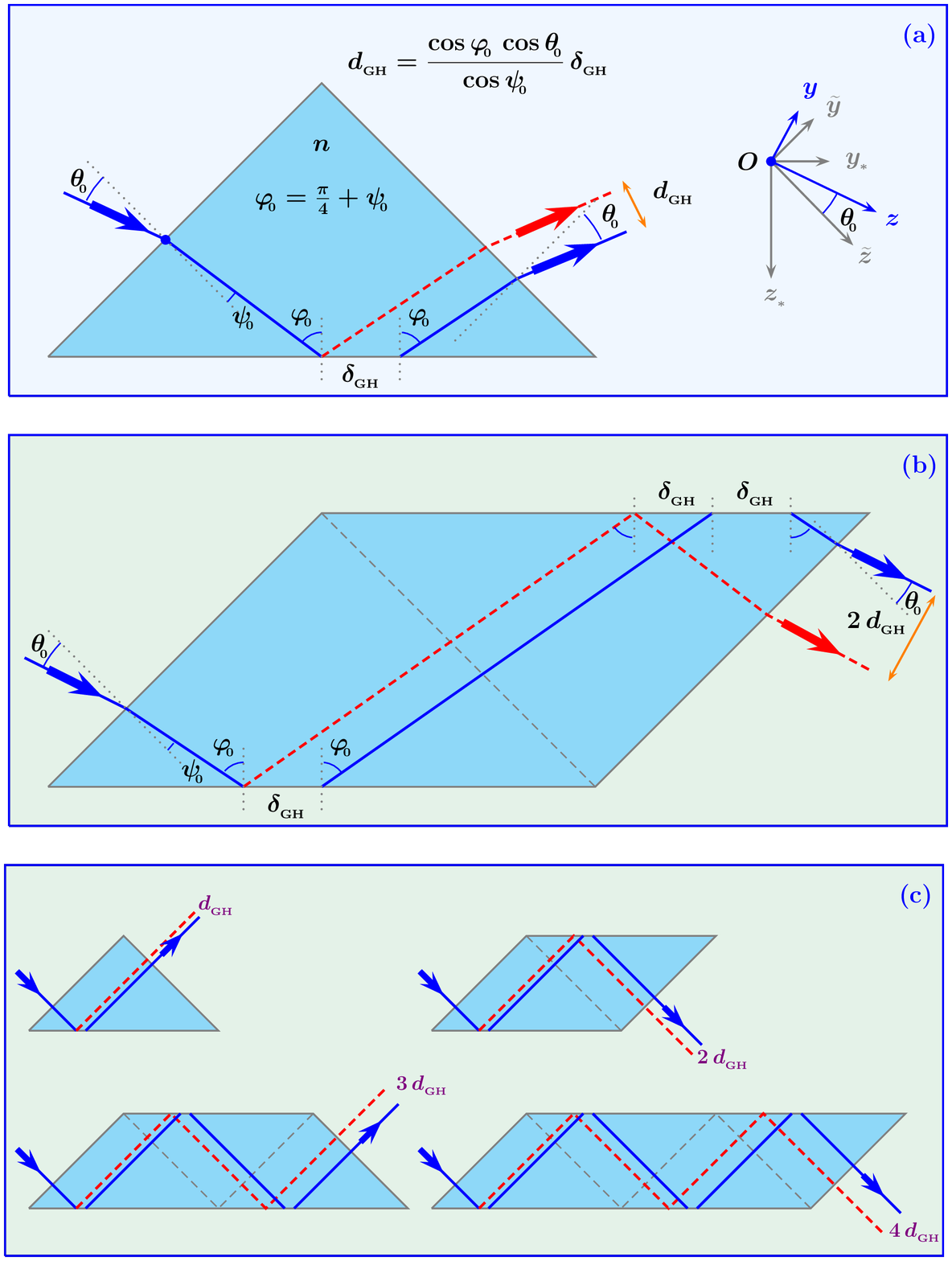}
		\vspace{-1.5cm}
		\caption{{\bf Dielectric blocks geometry.} Dielectric structure composed by $N$ triangular prisms. The GH shift of the outgoing beam with respect to the path predicted by geometrical optics is proportional to the number of triangular prisms, $N\,d_{_{\rm GH}}$.}
\label{fig1}
\end{figure}

\newpage

\begin{figure}[ht]
	\vspace{-2.2cm}
		\hspace{-1.3cm}
		\includegraphics[height=24cm, width=18cm]{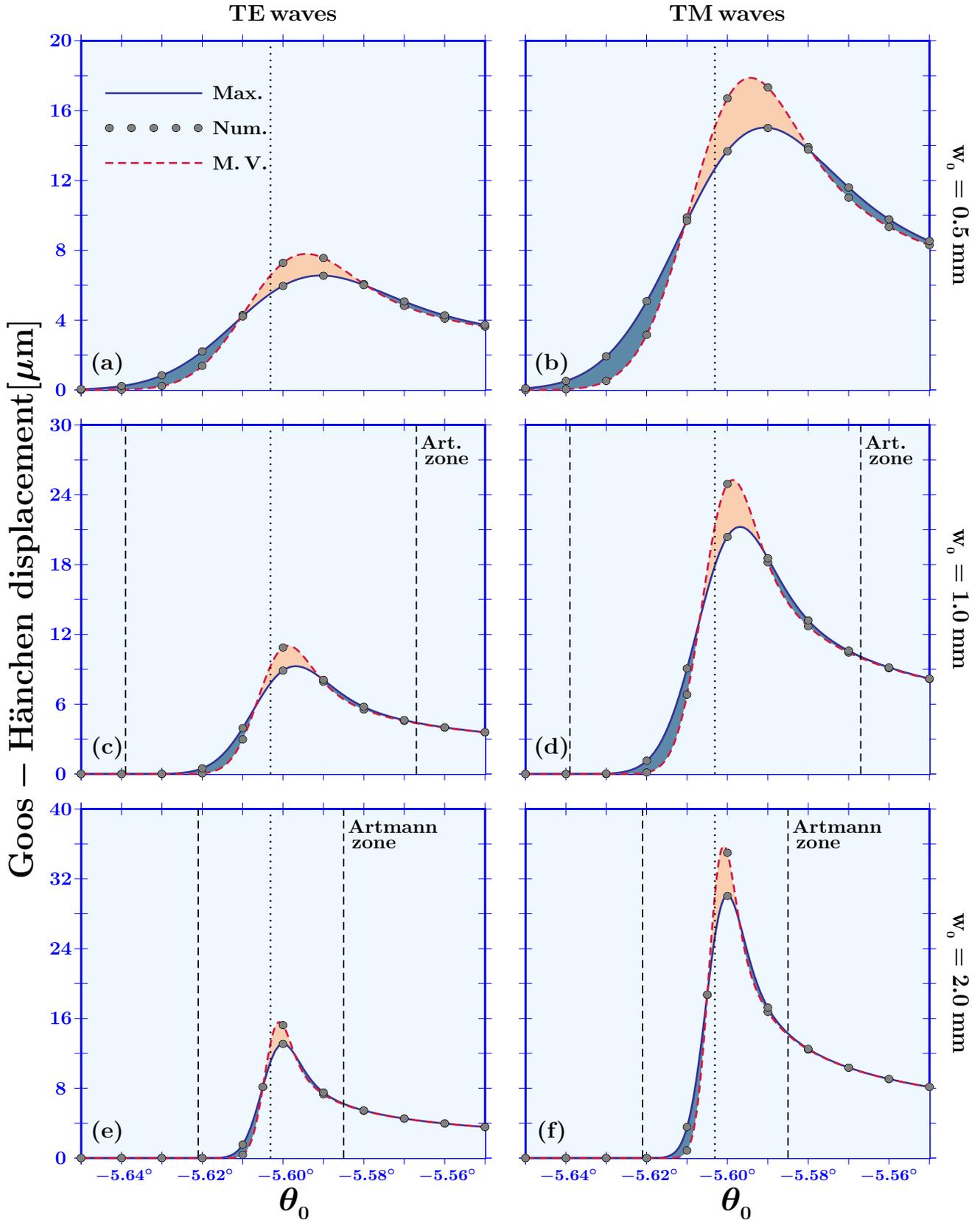}
		\vspace{-1.5cm}
		\caption{{\bf GH shift for Gaussian optical beams.} The GH shift of the intensity maximum and mean value is plotted, for TE and TM waves, as a function of the incidence angle $\theta_{\0}$ for  Gaussian beams with  $\lambda=0.633\,{\rm nm}$ and different beam waist ${\rm w}_{\0}=0.5,\,1.0\,,2.0\,\,{\rm mm}$.  The analytical expressions for the maximum, given in Eqs.\,(\ref{newf},\ref{newfTM}) and represented by solid blue lines, and for the mean value, given in  Eqs.\,(\ref{newf2},\ref{newf2TM}) and represented by dashed red lines, are in excellent agreement with the numerical data (gray dots). The difference between these curves clearly shows the breaking of symmetry in the critical region, $\theta_{\0}\in [\,\theta_{_{\rm cri}}-\frac{\lambda}{{\rm w}_{\0}}\,,\,\theta_{_{\rm cri}}+\frac{\lambda}{{\rm w}_{\0}}\,]$.}
\label{fig2}
\end{figure}

\newpage

\begin{figure}[ht]
	\vspace{-2.2cm}
		\hspace{-1.3cm}
		\includegraphics[height=24cm, width=18cm]{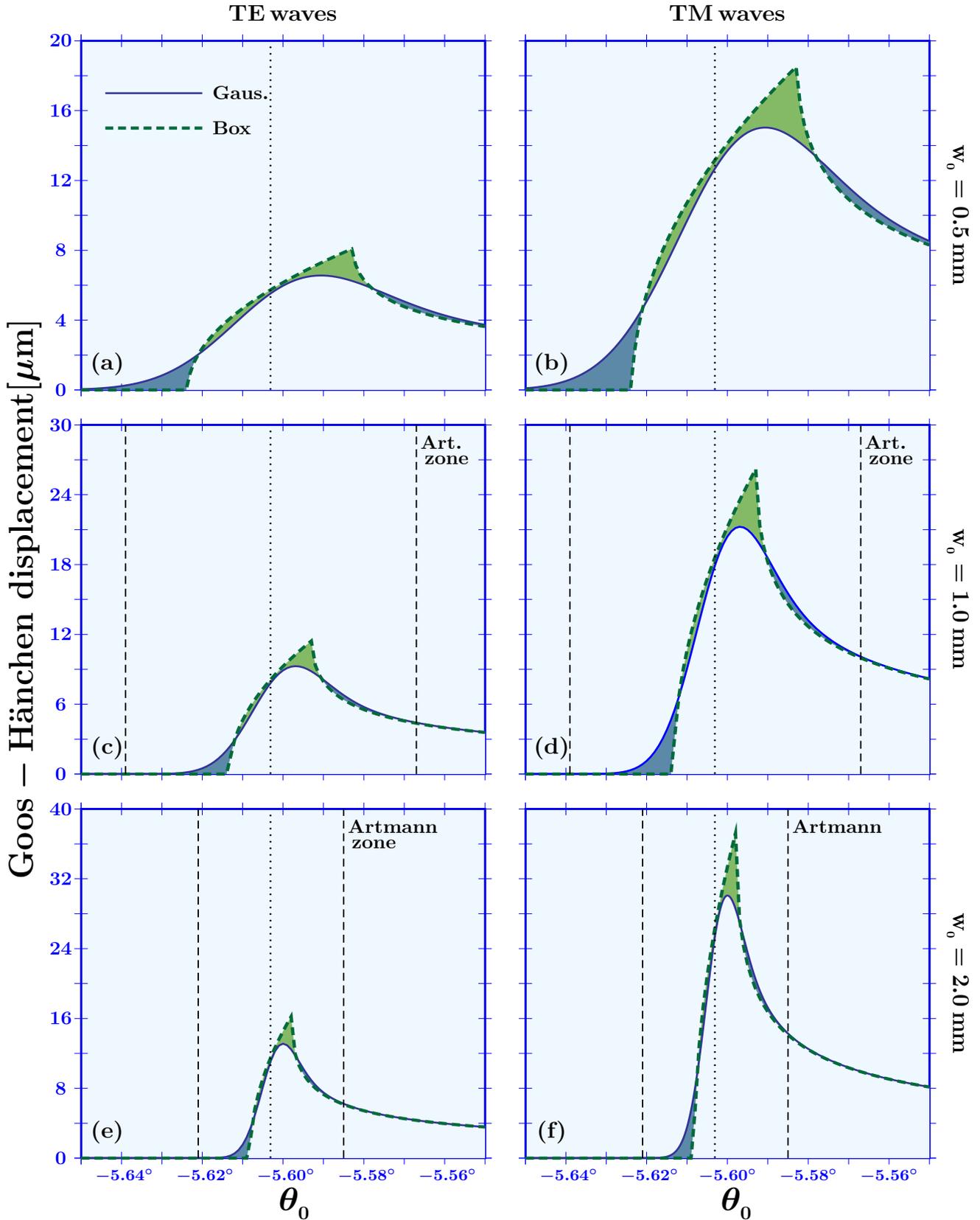}
		\vspace{-1.5cm}
		\caption{{\bf Dependence of the GH shift on the angular distribution shape.} In the critical region,  $\theta_{\0}\in [\,\theta_{_{\rm cri}}-\frac{\lambda}{{\rm w}_{\0}}\,,\,\theta_{_{\rm cri}}+\frac{\lambda}{{\rm w}_{\0}}\,]$,
the GH shift is dependent on the shape of the angular distribution. As an example, we plot the intensity maximum for gaussian (solid blue lines) and box (dashed green lines) angular distributions. In the Artmann zone, where the plane wave limit is valid no difference is found.   }
\label{fig3}
\end{figure}

\end{document}